\documentclass[onecolumn,nofootinbib,preprintnumbers,longbibliography,10pt]{revtex4-1}

\usepackage{amssymb}
\usepackage{graphics}
\usepackage{graphicx}
\usepackage{amsmath}
\usepackage{amsfonts}
\usepackage{bm}

\newlength{\extraspace}
\newlength{\extraspaces}
\usepackage[figtopcap]{subfigure}

\newcommand{\be}{\begin{equation}
\addtolength{\abovedisplayskip}{\extraspaces}
\addtolength{\belowdisplayskip}{\extraspaces}
\addtolength{\abovedisplayshortskip}{\extraspace}
\addtolength{\belowdisplayshortskip}{\extraspace}}
\newcommand{\ee}{\end{equation}}

\newcommand{\ba}{\begin{eqnarray}
\addtolength{\abovedisplayskip}{\extraspaces}
\addtolength{\belowdisplayskip}{\extraspaces}
\addtolength{\abovedisplayshortskip}{\extraspace}
\addtolength{\belowdisplayshortskip}{\extraspace}}
\newcommand{\ea}{\end{eqnarray}}

\usepackage{color,soul}
\usepackage[colorlinks,citecolor=blue,urlcolor=blue,linkcolor=blue]{hyperref}

\begin{document}

\title{ Black holes solutions in power-law Maxwell-$f(T)$ gravity in diverse dimensions}

 \author{G.G.L. Nashed}
\email{nashed@bue.edu.eg}
\affiliation {Centre for Theoretical Physics, The British University in Egypt, P.O. Box 43, El Sherouk City, Cairo 11837, Egypt\\
Int.\; Lab.\;  Theor. Cosmology, Tomsk State University of Control Systems and Radioelectronics (TUSUR), 634050 Tomsk, Russia}
\author{Kazuharu Bamba}
\email{bamba@sss.fukushima-u.ac.jp}
\affiliation{Division of Human Support System, Faculty of Symbiotic Systems Science, Fukushima University, Fukushima 960-1296, Japan}

\begin{abstract}
We investigate the solutions of black holes in $f(T)$ gravity with nonlinear power-law Maxwell field, where $T$ is the torsion scalar in teleparalelism. In particular, we introduce the Langranian with diverse dimensions in which the quadratic polynomial form of $f(T)$ couples with the nonlinear power-law Maxwell field. We explore the leverage of the nonlinear electrodynamics on the space-time behavior. It is found that these new black hole solutions tend towards those in general relativity without any limit. Furthermore, it is demonstrated that the singularity of the curvature invariant and the torsion scalar is softer than the quadratic form of the charged field equations in $f(T)$ gravity and much milder than that in the classical general relativity because of the nonlinearity of the Maxwell field. In addition, from the analyses of physical and thermodynamic quantities of the mass, charge and the Hawking temperature of black holes, it is shown that the power-law parameter affects the asymptotic behavior of the radial coordinate of the charged terms, and that a higher-order nonlinear power-law Maxwell field imparts the black holes with the local stability.
\end{abstract}

\maketitle

\section{Introduction}\label{S1}

Over the past years, black holes (BHs) possessing linear and nonlinear Maxwell fields have attracted tremendous attention. The study of charged BH solutions is relevant because BH created in a collider may generally possess an electric field. Moreover, the difference between the electroweak and Planck scales is still an unsolved challenge that is known as the hierarchy problem. This problem has been tackled in the frame of theories with extra spatial dimensions. Further, this work is an attempt at elucidating higher-dimension charged BH solutions. The first higher-dimension spherically symmetric BH solution was derived in  \cite{1963NCim...27..636T}, after which it was generalized in  \cite{1986AnPhy.172..304M}. Moreover, a higher-dimension charged BH solution, which is a generalization of the Reissner-Nordstr\"om BH, was derived in \cite{0256-307X-8-9-001}. There are different sets of the charged BH solutions, which were derived by the Brans-Dicke and Lovelock theories; their physical properties have been studied   \cite{Dehghani:2006ke,Dehghani:2006zi,Dehghani:2008qr,Hendi:2010bk}.  The research on the higher dimensions of the Kerr-Newman solution is in progress, although the slowly rotating BH solution has been derived in  \cite{Mignemi:1992pm,ElHanafy:2015egm,Volkov:1997qb,Ghosh:2007jb,Kim:2007iw,Sheykhi:2008rm}.

It is known that the Maxwell theory is invariant under conformal transformation in four dimensions, although it is not invariant in a higher dimension. The lack of this transformation has been explicitly studied in higher dimensions employing the nonlinear power-law Maxwell field  \cite{Hassaine:2007py,Hassaine:2008pw}. The conformal invariance in a higher dimension was considered to derive a similar four-dimensional Reissner-Nordstr\"om BH with extra dimensions. Thus, the present study is aimed at deriving diverse-dimensional charged BH solutions within the frame of the modified teleparallel equivalent of general relativity (TEGR), namely, { f(T)-gravitational theory}.

There are various reasons, ranging from our accelerated expansion of the universe and its dark energy in the astrophysical tests, for researchers to consider modifying general relativity (GR). Among the modifications, the Brans-Dicke  \cite{1961PhRv..124..925B,1962PhRv..125.2163D}, Lovelock \cite{1971JMP....12..498L}, $f(R)$ \cite{1970MNRAS.150....1B,Sotiriou:2008rp,Nojiri:2017ncd}, and $f(T)$ (refer to \cite{Cai:2015emx} for more details) gravitational theories have gained much attraction for different reasons. In the present study, we concentrate on the $f(T)$ gravity for many reasons, such as the fact that the Lagrangian of this theory which depends on the torsion scalar only, which makes it easy to handle compared with other modified gravitational theories ($f(R)$  \cite{Cembranos:2012fd,Nojiri:2010wj}). Another major reason for focusing on $f(T)$ is the point that the gravitational field equations are of the second order, unlike those in the other modified theories  \cite{FF07,Awad:2017sau,1010.1041}.

TEGR is a theory, which was developed by Einstein to unify the gravitational and electromagnetic fields  \cite{Nashed:2005kn,Nashed:2008ys,Nashed:2009hn,Nashed:2007cu,2005physics...3046U,Nashed:2006yw}. The TEGR theory could be applied to calculate the conserved quantities, mass and angular momentum employing the energy-momentum tensor  \cite{Maluf:1995re,Maluf:2002zc}. The main motivation of the modification of TEGR theory was the issues recently appeared in observations that TEGR cannot explain \cite{Ferraro:2006jd}. $f(T)$, which exhibits many viable applications, is the modification of the TEGR theory \cite{Capozziello:2015rda,Nashed:2001im,Iorio:2012cm,2012ChPhL..29e0402G,Bamba:2014zra}. By employing this theory, not only inflation \cite{Ferraro:2006jd} in the early universe but also the late-time cosmic acceleration could be explained \cite{Bengochea:2008gz,2010PhRvD..81l7301L,Wu:2010av,2011JCAP...01..009D,Awad:2017sau,2011JCAP...01..021B,2010arXiv1008.4036B,2012PhRvD..85j4036B,Aviles:2013nga,Jamil:2012vb,Ferraro:2011us,Ferraro:2011zb,Sebastiani:2010kv,
 Salako:2013gka,Haghani:2012bt,Haghani:2013pea}. There exist a number of interesting applications of $f(T)$ gravity to the realm of cosmology \cite{Bamba:2013fta,Bamba:2012ka,Bamba:2012yr,Bamba:2013jqa,Bengochea:2008gz,2010PhRvD..81l7301L,Geng:2011aj,Otalora:2013tba,2013Ap&SS.344..269C,
 2011JCAP...01..009D,
Yang:2010hw,Bamba:2010wb,Capozziello:2011hj,Awad:2017ign,2011PhRvD..84d3527C,2012JCAP...01..002G,2012Ap&SS.338..195F,2012PhRvD..85l4007C,Bahamonde:2015zma}, as well as in the domain of astrophysics \cite{Capozziello:2012zj,Paliathanasis:2014iva,Gonzalez:2011dr,Bohmer:2011si,Nashed:2013bfa,Ruggiero:2015oka}. In the astrophysical domain, $f(T)$ possesses a new exact charged BH solution, which involves, in addition to the monopole term, a quadruple term whose contribution accrues from the quadratic $f(T)$ form, i.e., $T^2$. Based on the achievements recorded in $f(T)$ gravity, we explore the implication of the nonlinear power-law Maxwell field. Particularly, we derive the field equations of $f(T)$ coupled with the nonlinear power-law Maxwell field and apply their quadratic form, that is, $f(T)=T_0+\alpha T-\beta T^2$, to diverse-dimensional flat transverse sections.

The arrangements of the study are the followings. In Sec. \ref{S1}, we explain $f(T)$ gravity and show the field equations with the Maxwell field. In Sec. \ref{AdsSection}, we investigate the solutions of the charged static BH with the Anti-de-Sitter/de-Sitter (AdS/dS) behavior. We analyze the singularity and horizon structure of these BH solutions and calculate their energies in Sec. \ref{AdsSection}.
In Sec. \ref{S5}, we explore the charged AdS solutions of the rotating BH in power-law Maxwell-$f(T)$ gravity. In Sec. \ref{S7}, we calculate different thermodynamical quantities and establish the local stability of our BH solutions. Finally, the conclusions and discussions are given in Sec. \ref{S8}.

\newpage

\section{Basic formulations of $f(T)$ gravity}\label{S1}

The {f(T)-gravity} is a kind of extension of TEGR. In this theory, it is suitable to apply the vielbein (tetrads) fields, $e_i^\mu$ , as dynamic variables (the Greek indices run for the coordinate space and the Latin one spans for the tangent one) that constitute orthonormal basis of the tangent space at the spacetime each point. The relations between the covariant and contravariant tetrads and between the tetrads and metrics are represented by the following equation:
\begin{equation}\label{q3}
e^i_\mu e_i^\nu:={\delta_\mu}^\nu, \qquad \qquad e^i_\mu e_j^\mu:={\delta_j}^i, \qquad \qquad
 {\it g_{\mu \nu} :=  \eta_{i j} {e^i}_\mu {e^j}_\nu,} \qquad \qquad \eta_{i j}:= {e_i}^\mu {e_j}^\nu g_{\mu \nu},
\end{equation}
 where $\eta_{i j}=(+,-,-,- \cdots)$ is the $d$-dimensional Minkowskian metric of the tangent space. Unlike the symmetric Levi-Civita connection in GR, the (non-symmetric) Weitzenb\"ock connection in TEGR is defined as \cite{Wr}
\begin{equation}\label{wet} \overset{{\bf{w}}}{\Gamma}^\lambda_{\mu \nu} := {e_i}^\lambda~
\partial_\nu e^{i}_\mu.
\end{equation}
It follows from Eq. (\ref{wet}) that the torsion tensor is given by
\begin{equation}
\label{torsten}
{T^\alpha}_{\mu \nu} :=
\overset{{\bf{w}}}{\Gamma}^\alpha_{ \nu\mu }-\overset{{\bf{w}}}{\Gamma}^\alpha_{\mu \nu}
={e_i}^\alpha
\left(\partial_\mu{e^i}_\nu-\partial_\nu{e^i}_\mu\right).
\end{equation}
This tensor encodes all the information of the gravitational field.  The difference between the Levi-Civita and the Weitzenb\"{o}ck connections defines the contorsion tensor, which  is expressed by Eq. (\ref{torsten1}):
 \begin{equation}
\label{torsten1}
{K^\lambda}_{\mu \nu} :=\overset{{\bf{w}}}{\Gamma}^\lambda_{\mu \nu}-\overset{{\bf{\circ}}}{\Gamma}^\lambda_{\mu \nu},\end{equation} where $\overset{{\bf{\circ}}}{\Gamma}^\lambda_{\mu \nu}$ is the symmetric Levi-Civita connection. The super-potential can be defined by the foregoing equations, as follows:
\begin{equation}
\label{sup}
{S_\lambda}^{\mu \nu} :={K^{\mu \nu}}_\lambda+\delta^\mu_\lambda{T^{\alpha \nu}}_\alpha-\delta^\nu_\lambda{T^{\alpha \mu}}_\alpha, \end{equation}
{that has a skew symmetry in the last  two indices.} The torsion scalar with the following form can be defined by Eqs. (\ref{torsten}) and (\ref{sup}):
\begin{equation}
\label{Tor_sc}
T=\frac{1}{2}{S_\lambda}^{\mu \nu}{T^\lambda}_{\mu \nu}\equiv\frac{1}{4}
T^{\rho \mu \nu}
T_{\rho \mu \nu}
+\frac{1}{2}T^{\rho \mu \nu }T_{\nu \mu\rho }
-T_{\rho \mu }^{\ \ \rho }T_{\
\ \ \nu }^{\nu \mu }.
\end{equation}
The Lagrangian of TEGR mainly depends on the torsion scalar $T$ and the variation of the the Lagrangian with respect to the vierbeins can lead to the same field equations in GR.

With the same spirit of $f(R)$ gravity, the Lagrangian of TEGR described by $T$ can be expanded to an arbitrary function $f$ of $T$ as $f(T)$ as follows \cite{Cai:2015emx}:
\begin{equation}\label{q7a}
{\cal L}=\frac{1}{2\kappa}\int |e|f(T)~d^{d}x .
\end{equation}
Here, $|e|=\sqrt{-g}=\det\left({e^a}_\mu\right)$ is the determinant of the tetrad. Moreover, $\kappa$ is a constant in the $d$-dimensions with the form $\kappa =2(d-3)\Omega_{d-1} G_d$, where $G_d$ the Newtonian constant in the $d$-dimensions and $\Omega_{d-1}$ is the volume of the unit sphere in the $(d - 1)$-dimensions, given by
\begin{equation}
 \Omega_{d-1} = \frac{2\pi^{(d-1)/2}}{\Gamma[(d-1)/2]},
 \end{equation}
with $\Gamma$ a $\Gamma$-function.

 The action (\ref{q7a}) coupled with the power law-Maxwell Lagrangian is represented as
\begin{equation} \label{lag}
{\cal L}=\frac{1}{2\kappa}\int |e|f(T)~d^{d}x+\int |e|{\cal L}_{ em}~d^{d}x,
\end{equation}
where  ${\cal L}_{em}= {\cal F}^s$, with ${\cal F} = dQ$ and  $Q=Q_{\mu}dx^\mu$  is the
 electromagnetic potential 1-form \cite{Capozziello:2012zj} and $s$ is
a power-law parameter in terms of the Maxwell field.
When the power for the Maxwell field is equal to unity ($s=1$), the Lagrangian describes the ordinary Maxwell theory \cite{Nashed:2013bfa}.

By varing the action in Eq. (\ref{lag}) in terms of the tetrads, we acquire \cite{Cai:2015emx}:
\begin{eqnarray}\label{q8a}
& & I^\nu{}_\mu={S_\mu}^{\rho \nu} \partial_{\rho} T
f_{T T}+\left[e^{-1}{e^i}_\mu\partial_\rho\left(e{e_i}^\alpha
{S_\alpha}^{\rho \nu}\right)-{T^\alpha}_{\lambda \mu}{S_\alpha}^{\nu \lambda}\right]f_T
-\frac{1}{4} f \delta^\nu_\mu +\frac{1}{2}\kappa{{{\mathrm{
T}}^{{}^{{}^{^{}{\!\!\!\!\scriptstyle{em}}}}}}}^\nu_\mu \equiv0.
\end{eqnarray}
Here, we show $f(T)$ as $f$, and we describe
$f_{T}:= \frac{\partial f(T)}{\partial T}$ and
$f_{TT}:= \frac{\partial^2 f(T)}{\partial T^2}$.
Moreover, ${{{\mathrm T}^{{}^{{}^{^{}{\!\!\!\!\scriptstyle{em}}}}}}}^\nu_\mu$   means the energy-momentum tensor of the power law-Maxwell field defined by
\begin{equation}\label{en}
{{{\mathrm
T}^{{}^{{}^{^{}{\!\!\!\!\scriptstyle{em}}}}}}}^\nu_\mu=s{\cal F}_{\mu \alpha}{\cal F}^{\nu
\alpha}{\cal F}^{s-1}-\frac{1}{4} \delta_\mu{}^\nu {\cal F}^s\,,\end{equation} where ${\cal F}={\cal F}_{\mu \nu}{\cal F}^{\mu \nu}$.

Furthermore, the variation of Eq. (\ref{lag}) with respect to $\boldsymbol{Q_{\mu}}$,
which is the 1-form of the gauge potential, yields
\begin{eqnarray}\label{q8b}
&&\partial_\alpha\left( \sqrt{-g} {\cal F}^{\mu \nu}{\cal F}^{s-1} \right)=0\; .
\end{eqnarray}
It follows from Eq. (\ref{q8b}) clearly that when the power law $s=1$ the energy-momentum tensor of Eq. (\ref{q8b}) coincides with the linear form of Maxwell field \cite{2013PhRvD..88j4034N}.  Eq. (\ref{q8b}) determines the power law of {Maxwell field} in arbitrary dimensions.

In addition, Eq. (\ref{q8a}) can take the following form:
\begin{equation} \partial_\nu \Biggl[e{S}^{a \rho \nu} f_T\Biggr]=\kappa e
{e^a}_\mu \Biggl[t^{\rho \mu}+{{{\mathrm
T}^{{}^{{}^{^{}{\!\!\!\!\scriptstyle{em}}}}}}}^{\rho \mu}\Biggr],\end{equation}
where $t^{\nu \mu}$ is the energy-momentum tensor of the gravitational configuration, given by
\begin{equation} t^{\nu
\mu}=\frac{1}{\kappa}\Biggl[4f_T {S^\alpha}^{\nu
\lambda}{T_{\alpha \lambda}}^{\mu}-g^{\nu \mu} f\Biggr].
\end{equation}
The anti-symmetry of the tensor ${S}^{a \nu \lambda}$ leads
\begin{equation}
\partial_\mu \partial_\nu\left[e{S}^{a \mu \nu} f_T\right]=0,
\end{equation}
from which we have
\begin{equation} \label{q88}
\partial_\mu\left[e\left(t^{a \mu}+{{{\mathrm
T}^{{}^{{}^{^{}{\!\!\!\!\scriptstyle{em}}}}}}}^{a
\mu}\right)\right]=0.
\end{equation}
Hence, from Eq. (\ref{q88}) we find
\begin{equation} \label{qbb}
\frac{d}{dt}\int_V d^{(d-1)}x \ e \ {e^a}_\mu \left(t^{0 \mu}+{{{\mathrm
T}^{{}^{{}^{^{}{\!\!\!\!\scriptstyle{em}}}}}}}^{0
\mu}\right)+ \oint_\Sigma \left[e \ {e^a}_\mu \ \left(t^{j
\mu}+{{{\mathrm
T}^{{}^{{}^{^{}{\!\!\!\!\scriptstyle{em}}}}}}}^{j
\mu}\right)\right]=0.
\end{equation}
Equation (\ref{qbb}) denotes the conservation law of ${{{\mathrm T}^{{}^{{}^{^{}{\!\!\!\!\scriptstyle{em}}}}}}}^{\lambda \mu}$ as well as the pseudo tensor $t^{\lambda \mu}$, which describes the energy-momentum tensor of gravitation in $f(T)$ gravity \cite{Ulhoa:2013gca}.
Thus, the energy-momentum tensor of $f(T)$ gravity in the ($d-1$)--dimensions contained in the volume $V$ reads
\begin{equation} \label{qbbb}
P^a=\int_V d^{(d-1)}x
\ e \ {e^a}_\mu \left(t^{0 \mu}+{{{\mathrm
T}^{{}^{{}^{^{}{\!\!\!\!\scriptstyle{em}}}}}}}^{0
\mu}\right)=\frac{1}{\kappa}\int_V d^{(d-1)}x  \partial_\nu\left[e{S}^{a 0
\nu} f_T\right],
\end{equation}
which corresponds to TEGR when $f(T)=T$ \cite{Maluf:2002zc}. The above equation (\ref{qbbb}) is the conserved four-momentum equation for any configuration that behaves as a flat spacetime. In this research, we derived a class of BH solutions, which behaves asymptotically as an AdS spacetime. Therefore, it was necessary to conduct the calculations of the conserved quantities concerning a pure AdS/dS space to avoid the conserved quantities having infinite value because the asymptotic behavior of the solutions of BH is similar to that of AdS. With the difference of the energy of the pure AdS BH solution from that of the AdS space, the total energy of the AdS BH could be found. Hence, for the calculation of the conserved quantities, we subtract the effects of the AdS space, which we describe by using the subscription ``$r$'' to regularized value. From Eq. (\ref{qbb}), we get
\begin{equation} \label{conr}
\frac{d}{dt}\int_V d^{(d-1)}x \ e \ {e^a}_\mu \left(t_{r}^{0 \mu}+{{{\mathrm
T}^{{}^{{}^{^{}{\!\!\!\!\scriptstyle{em}}}}}}}^{0
\mu}\right)+ \oint_\Sigma \left[e \ {e^a}_\mu \ \left(t_{r}^{j
\mu}+{{{{\mathrm
T}_r^{{}^{{}^{^{}{\!\!\!\!\scriptstyle{em}}}}}}}}{}^{j
\mu}\right)\right]=0.\end{equation}  Thus Eq. (\ref{conr}) is the regularized conservation law of any spacetime that behaves as (A)dS.

\section{New Anti-de-Sitter solutions of black holes in power law Maxwell-$f(T)$ gravity}
\label{AdsSection}

{ In this section, we will derive the AdS charged BH solutions in $d$--dimensions for the power-law  Maxwell-$f(T)$-gravity. For this aim, we will use the flat transverse sections in the  $d$--dimensions ($t$, $r$,
$\theta_1$, $\theta_2$,
$\cdots$, $\theta_{i}$, $z_1$, $z_2$ $\cdots$ $z_k$),
where $k=1,2 \cdots$ $d-i-2$, $0\leq r< \infty$, $-\infty < t < \infty$, $0\leq \theta_{i}< 2\pi$, $-\infty < z_k < \infty$, with the following { vielbein} form  \cite{Capozziello:2012zj,Awad:2017tyz}:
\begin{equation}\label{tetrad}
\hspace{-0.3cm}\begin{tabular}{l}
   $\left({e^{i}}_{\mu}\right)=\left( \sqrt{N(r)}, \; \frac{1}{\sqrt{N(r)g(r)}}, \; r, \;
r, \; r\;\cdots \right)$.
\end{tabular}
\end{equation}
Here, $t$ and $r$ are time and the radial coordinate, respectively.
The metric associated of the {vielbein} in Eq.~(\ref{tetrad}) assumes the following form:}
\begin{equation}
\label{m2}
ds^2=
N(r)dt^2-\frac{1}{N(r)g(r)}dr^2-r^2\left(\sum_{i=1}^{n}d\theta^2_i+\sum_{k=1}^{d-n-2}
dz_k^2\right),
\end{equation}
with $N(r)$ and $g(r)$ two unknown functions $r$\footnote{{In 4-dimension the metric (\ref{m2}) yields \[ds^2=
N(r)dt^2-\frac{1}{N(r)g(r)}dr^2-r^2\gamma_{ij}dx^idx^j \,.\] where $\gamma_{ij}dx^idx^j$
represents the line element of a two-dimensional surface with
constant curvature $k = -1, 0, 1$, and the indices $(i, j) = 1, 2$. The well-known solutions of GR, such as the Schwarzschild and the
Reissner-Nordstr¨om geometries, correspond to spherical horizon structure where $k = 1$. In this work, however, we shall consider solutions with a flat
horizon structure where $k = 0.$}}. By combining the tetrad in Eq.~(\ref{tetrad}) into the scalar torsion $T$ in Eq.~(\ref{Tor_sc}), we obtain
\begin{equation}\label{df}
T=\frac{(d-2)g}{r}\Big[N'+\frac{(d-3)N}{r}\Big],
\end{equation}
where $N'=\frac{dN(r)}{dr}$ and $g'=\frac{dg(r)}{dr}$.
In the following, we omit the arguments of $N(r)$, $g(r)$, $N'(r)$ and $g'(r)$.  Owing to the success of the power-law form of $(T)$ to describe cosmology \cite{Nesseris:2013jea,Nunes:2016qyp,Basilakos:2018arq}, we focus on the quadratic form
\begin{equation}\label{powellaw}
 f(T)=T_0+\alpha T-\beta T^2,
\end{equation}
where $T_0$, $\alpha$ and $\beta$ are constants.

\subsection{Asymptotically static AdS BHs with the power-law Maxwell field}\label{S2}

Regarding the vanishing of the electromagnetic sector, i.e., ${{{\mathrm
T}^{{}^{{}^{^{}{\!\!\!\!\scriptstyle{em}}}}}}}^\nu_\mu=0$, the results are identical to those obtained in Ref.~\cite{Capozziello:2012zj,Awad:2017tyz, 2018arXiv181103658N}. However, novel results can be found for the non-vanishing of the electromagnetic field, and this could be explained by the tetrads in Eq.~(\ref{tetrad}) in the field equations (\ref{q8a}) and (\ref{q8b}) by utilizing the vector potential 1-form\footnote{{Through out the rest of this study we will put $\kappa=1$.}}.
\begin{equation}\label{pot}
Q(r) = \phi(r)dt.
\end{equation}
The non-zero components of the field equations read
\begin{eqnarray} \label{dfc}
& & I^r{}_r=2Tf_T-f-(2s-1)\Big\{-2\phi'^2g(r)\Big\}^s=0,\nonumber\\
& & I^{z_1}{}_{z_1}= I^{z_2}{}_{z_2}=\cdots \cdots =I^{z_{d-n-2}}{}_{z_{d-n-2}}=  \frac{f_{TT} [r^2T+(d-2)(d-3)N]gT'}{r(d-2)}+\frac{f_T}{2r^2}\Biggl\{2r^2gN''+2(3d-8)rgN'\nonumber\\
& & +4(d-3)^2Ng+2(d-3)rNg'+r^2N'g'\Biggr\}-f+\Big\{-2\phi'^2g(r)\Big\}^s=0, \nonumber\\
& & I^t{}_t= \frac{2(d-2)Ngf_{TT} T'}{r}+\frac{(d-2)f_T[2\{(d-3)Ng+rgN'\}+rNg']}{r^2}-f-(2s-1)\Big\{-2\phi'^2g(r)\Big\}^s=0,\nonumber\\
& &
\end{eqnarray}
where $\phi'=\frac{d\phi}{dr}$. Various observations including the followings are extracted from Eq.~(\ref{dfc}):\vspace{0.1cm}\\
(i)-Equation (\ref{dfc}) is reduced to those derived in \cite{Awad:2017tyz} for $s=1$.\vspace{0.1cm}\\
(ii)-When $s=1/2$, the above system does not yield any solution because the charged terms of Eq.~(\ref{dfc}) are imaginary. Hence, the case (ii) is excluded in this study.

To derive an exact and well-behaved physical solution for the above system, we considered $T_0=-\frac{1}{12\beta}$ and $\alpha=1$ and assumed that $s$ possessed odd values. Other factors prevented the good behavior of the solution and imparted it with an imaginary value. By employing the pervious constraints, we acquire the following general solution of the above differential equations in $d$-dimensions
\begin{eqnarray} \label{sol}
& &  N(r)=\frac{r^2}{6(d-1)(d-2)\beta}+\frac{c_1}{r^{d-3}}
+\frac{3\;2{^{^s}}(2s-1)^2c_2{}^{2s}}{2(d-2)(2s+1-d)r^{\frac{2[1+(d-4)s]}{2s-1}}}+
\frac{2^{^{\frac{3s-1}{2}}}(2s-1)^{\frac{5}{2}}c_2{}^{3s}\sqrt{6\beta}}{(d-2)[(d-1)+(d-4)s]r^{\frac{[2+(3d-10)s]}{2s-1}}},\nonumber\\
& &
 g(r)=\frac{1}{\Bigg[c_2{}^{s}\sqrt{\frac{2{^{^{(s-1)}}}\; (-1)^{^{1+s}}\; 6\;\beta(2s-1)}{r^{^{\frac{2s(d-2)}{2s-1}}}}}+1\Bigg]^2},\qquad \qquad {\cal F}_{tr}=\frac{c_2}{r^{^{\frac{(d-2)}{2s-1}}}}+
 \frac{c_2{}^{s+1}\sqrt{2{^{^{(s-1)}}}\;(-1)^{^{1+s}}\;6\;\beta(2s-1)}}{{r^{^{\frac{(s+1)(d-2)}{2s-1}}}}},\nonumber\\
& &
 \phi(r)=\frac{c_2(2s-1)}{(d-1-2s){r^{^{\frac{(d-1-2s)}{2s-1}}}}}+
 \frac{c_2{}^{s+1}\sqrt{2{^{^{(s-1)}}}\;(-1)^{^{1+s}}\;6\;\beta(2s-1)^{5/2}}}{{[d-1+s(d-4)]r^{^{\frac{[d-1+s(d-4)]}{2s-1}}}}},
\end{eqnarray}
where $s$ takes a odd value, ${\cal F}_{tr}$ is the electric field, and $\phi(r)$ is the gauge potential 1-form. Equation (\ref{sol}) clearly expresses that $s$  must not be equal to half ($s\neq 1/2$). Additionally, Eq. (\ref{sol}) is a generalization of the one presented in \cite{Awad:2017tyz} and is reduced to them at $s=1$. For an asymptotic AdS/dS spacetime, we have
 \begin{equation}\label{lam}    \Lambda_\mathrm{eff}=\frac{1}{6(d-1)(d-2)\beta}.
\end{equation}
{Equation (\ref{lam}) ensures that  black hole solution (\ref{sol}) has no corresponding in TEGR  upon taking
the limit $\beta \rightarrow 0$, which means this charged black hole solution has no analogue in GR or TEGR.}

By combining Eqs. (\ref{sol}) and (\ref{lam}), we get
\begin{eqnarray} \label{sol1}
 & &N(r)=\Lambda_\mathrm{eff}r^2-\frac{m}{r^{d-3}}
+\frac{3\;2{^{^s}}(2s-1)^2q^{2s}}{2(d-2)(2s+1-d)r^{\frac{2[1+(d-4)s]}{2s-1}}}+
\frac{2^{^{\frac{3s-1}{2}}}(2s-1)^{\frac{5}{2}}q^{3s}\sqrt{6\beta}}{(d-2)[(d-1)+(d-4)s]r^{\frac{[2+(3d-10)s]}{2s-1}}},\nonumber\\
& &
 g(r)=\frac{1}{\Bigg[q{}^{s}\sqrt{\frac{2{^{^{(s-1)}}}\; (-1)^{^{1+s}}\; 6\;\beta(2s-1)}{r^{^{\frac{2s(d-2)}{2s-1}}}}}+1\Bigg]^2},\qquad \qquad {\cal F}_{tr}=\frac{q}{r^{^{\frac{(d-2)}{2s-1}}}}+
 \frac{q{}^{s+1}\sqrt{2{^{^{(s-1)}}}\;(-1)^{^{1+s}}\;6\;\beta(2s-1)}}{{r^{^{\frac{(s+1)(d-2)}{2s-1}}}}},\nonumber\\
& &\phi(r)=\frac{q(2s-1)}{(d-1-2s){r^{^{\frac{(d-1-2s)}{2s-1}}}}}+
 \frac{q{}^{s+1}\sqrt{2{^{^{(s-1)}}}\;(-1)^{^{1+s}}\;6\;\beta(2s-1)^{5/2}}}{{[d-1+s(d-4)]r^{^{\frac{[d-1+s(d-4)]}{2s-1}}}}},
\end{eqnarray}
where $c_1=-m$, and $c_2=q$. It follows from Eq.~(\ref{sol1}) that a kind of cosmological constant can appear in $f(T)$ gravity \cite{Iorio:2012cm,Kofinas:2015hla}. Moreover, it is seen from Eq. (\ref{sol1}) that $s\leq\frac{d-1}{2}$ so that the  monopole term, first term of $\phi(r)$ in Eq.~(\ref{sol1}), could exhibit finite behavior. However, the second term of $\phi(r)$ in Eq.~(\ref{sol1}) is always finite at $r\rightarrow 0$.

To discuss some of the physical properties of the aforementioned BH solution, we substituted Eq.~(\ref{sol1}) into Eq.~(\ref{m2}) and obtained the metric spacetime in the following form:
\begin{eqnarray}
\label{metric}
  &&
  \!\!\!\!\!\!\!
  ds{}^2=\Biggl[\Lambda_{eff}r^2-\frac{m}{r^{d-3}}
+\frac{3\;2{^{^s}}(2s-1)^2q^{2s}}{2(d-2)(2s+1-d)r^{\beta_1}}+
\frac{2^{^{\frac{3s-1}{2}}}(2s-1)^{\frac{5}{2}}q^{3s}\sqrt{6\beta}}{(d-2)[(d-1)+(d-4)s]r^{\beta_2}}\Biggr]dt^2-r^2\left(\sum_{i=1}^{n}d\theta^2_i+\sum_{k=1}^{d-n-2}
dz_k^2\right)
\nonumber\\
& &\ \ \
-\frac{\!dr^2}{\Biggl[\Lambda_{eff}r^2-\frac{m}{r^{d-3}}
+\frac{3\;2{^{^s}}(2s-1)^2q^{2s}}{2(d-2)(2s+1-d)r^{\beta_1}}+
\frac{2^{^{\frac{3s-1}{2}}}(2s-1)^{\frac{5}{2}}q^{3s}\sqrt{6\beta}}{(d-2)[(d-1)+(d-4)s]r^{\beta_2}}\Biggr]\Bigg[q{}^{s}\sqrt{\frac{2{^{^{(s-1)}}}\; (-1)^{^{1+s}}\; 6\;\beta(2s-1)}{r^{\beta_3}}}+1\Bigg]^2},\!
\end{eqnarray}
where $\beta_1=\frac{2[1+(d-4)s]}{2s-1}$, $\beta_2=\frac{[2+(3d-10)s]}{2s-1}$ and $\beta_3={\frac{2s(d-2)}{2s-1}}$.
As expected, we have obtained a solution, which behaves asymptotically as (A)dS dS according to the sign of the dimensional parameter $\beta$ because the constants $\beta_1$, $\beta_2$ $\beta_3$ are always positive. Additionally,
 Eq.~(\ref{sol1}) is a generalization of the ones derived in Ref.~\cite{Capozziello:2012zj,Awad:2017tyz}, owing to the application of a more general power-law { Maxwell-$f(T)$-gravity}.

Furthermore, we investigate the singularity structure of the BH solution in Eq.~(\ref{sol1}),by calculating the curvature and torsion invariants. The curvature scalars were calculated from the metric in Eq.~(\ref{metric}, whereas the torsion scalar was calculated employing the vierbeins in Eq.~(\ref{tetrad}). By calculating the Ricci scalar, the Ricci tensor square, and the Kretschmann scalar, we find
\begin{eqnarray}
R\approx \frac{C_1(r)}{\sqrt{\beta}r^{^{\frac{{{s(d-2)}}}{2s-1}}}},
 \qquad R^{\mu \nu}R_{\mu \nu}\approx
\frac{C_2(r)}{\sqrt{\beta}r^{^{\frac{{{2s(d-2)}}}{2s-1}}}},
\qquad
K\approx  R^{\mu \nu \lambda \rho}R_{\mu \nu \lambda \rho}\approx
\frac{C_3(r)}{\sqrt{\beta}r^{^{\frac{{{2s(d-2)}}}{2s-1}}}},
\end{eqnarray}
and the torsion scalar exhibited the following form:
\begin{eqnarray} \label{tors}
T(r)\approx\frac{C_4(r)}{\sqrt{\beta}r^{^{\frac{{{s(d-2)}}}{2s-1}}}},
\end{eqnarray}
where $C_i(r)$ is the lengthy polynomial function of $r$.
 The foregoing invariants clearly indicate that there is singularity at $r=0$.  At  $limit_{r=0}$, , the above invariants exhibited the following form $(K,R_{\mu
\nu}R^{\mu \nu}) \sim r^{^{\frac{{{-2s(d-2)}}}{2s-1}}}$  and  $(R,T)\sim r^{^{\frac{{{-s(d-2)}}}{2s-1}}}$   (dissimilar to the
BH solutions that were derived from the linear Maxwell-$f(T)$ in which $(K,R_{\mu
\nu}R^{\mu \nu}) \sim r^{^{{{{-2(d-2)}}}}}$ and  $(R,T)\sim r^{^{{{{-(d-2)}}}}}$. the asymptotic behaviors of the curvature and torsion invariants of the solution (Eq. \ref{sol1})  ) were different from those of the BH solution that were derived from the Einstein-Max- ell theory in the GR and TEGR formulations, which behaved as  $(K ,R_{\mu \nu}R^{\mu \nu})\sim
r^{-2d}$ and $(R,T) \sim  r^{-d}$. This clearly indicated that the singularity of the charged BH solution  (\ref{sol1}) was softer than that, which was obtained in the linear Maxwell-$f(T)$ and much softer than those of GR and TEGR for the charged case. Finally, it is noteworthy that although the solution  (\ref{sol1}) possessed different components, $g_{tt}$ and $g_{rr}$, its Killing vector and event horizons were equal.

\begin{figure}[ht]
\centering
\subfigure[~The $N(r)$ in 4-dim.]{\label{fig:pot1}\includegraphics[scale=0.3]{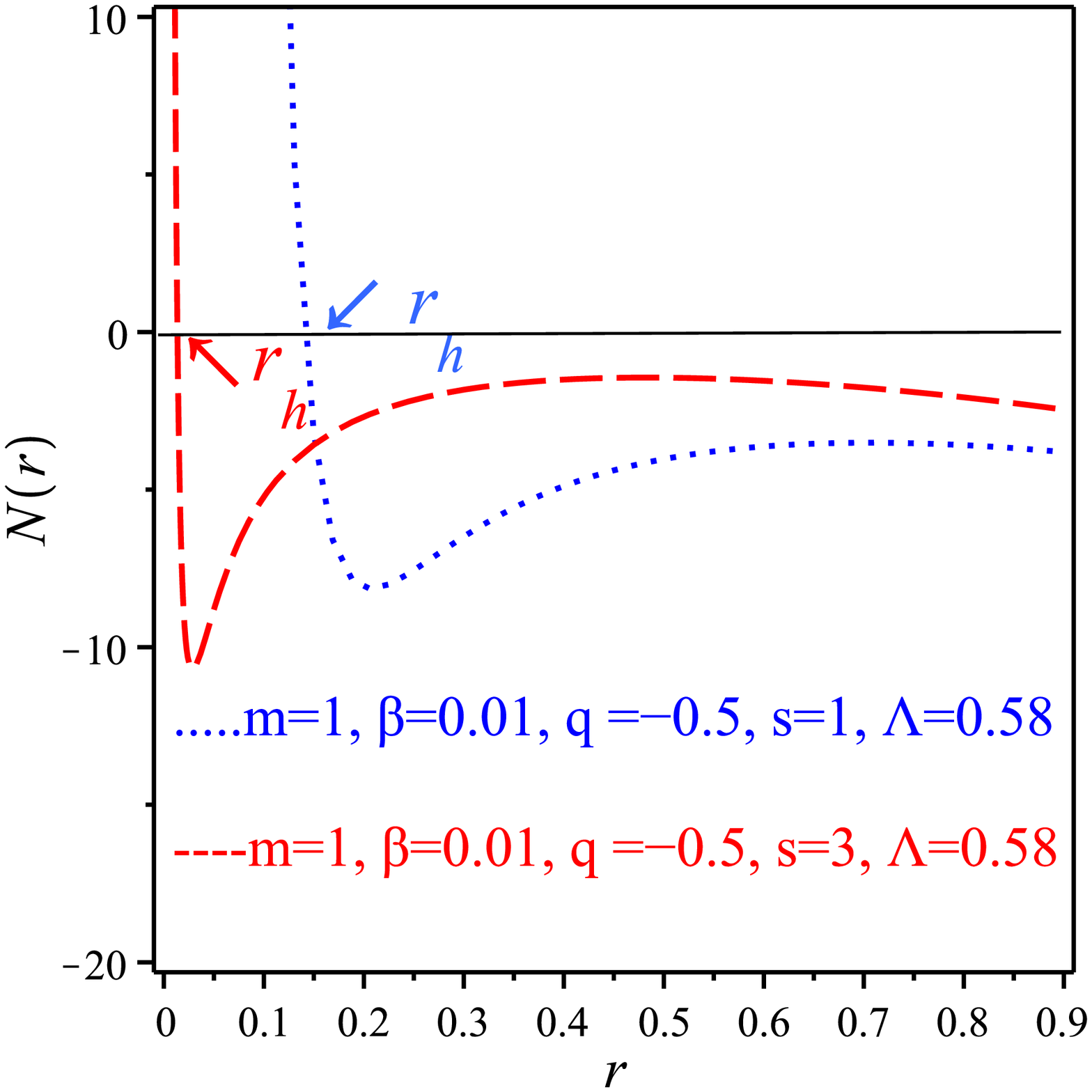}}
\subfigure[~The $N(r)$ in 5-dim.]{\label{fig:pot1}\includegraphics[scale=0.3]{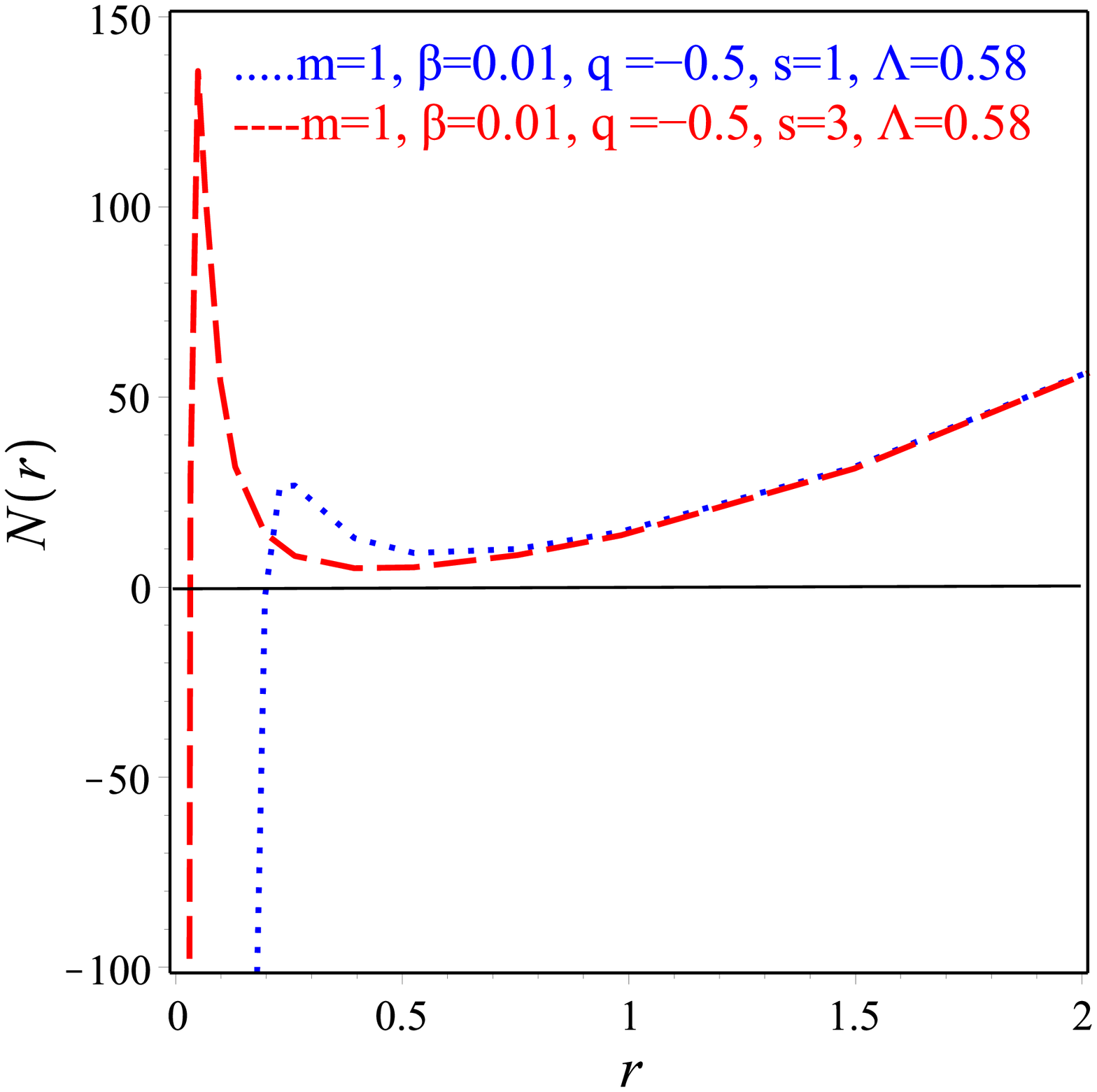}}
\caption{{\it{The  function $N(r)$ of solution (\ref{sol1}) of power law Maxwell-$f(T)$
gravity in four and five dimensions, for various values of  the power law parameter $s$. The term $r_h$ denotes the black hole of
inner Cauchy horizon.}}}
\label{Fig:1}
\end{figure}

To investigate the horizons of the solution  (\ref{sol1}), it was necessary to calculate the roots of the function $N(r) = 0$. {{We plotted the function $N(r)$  versus the radial coordinate, $r$, in four and five dimensions for the various values of the model parameters, as depicted in  figure}} \ref{Fig:1}.

{Furthermore, the aforementioned properties of the BH solution were explained differently by expressing the horizon mass–radius,  $m_h$, of  $m$  which corresponds to  $r_h$, that representing  the global properties of a horizon as obtained by setting   $N(r_h)=0$, namely}
\begin{equation} \label{hor1}
{m_h}=\frac{r_h^{(d-3)}\left(36\beta(2s-1)^2q^{2s}2^s\mathrm{A}+(d-1-2s)\Big[24\;2^s\beta q^{3s}(2s-1)^{5/2}\sqrt{6\beta}\;r_h{}^{\frac{-(d-2)s}{2s-1}}-r_h{}^{\frac{2[2+(d-4)s]}{2s-1}}(d-2)\mathrm{A}\Big]\right)}{24r_h{}^{\frac{2[1+(d-4)s]}{2s-1}} \beta (d-2)(d-1-2s)\mathrm{A}},
\end{equation}
where $\mathrm{A}=(1+s)d-1-4s$. Noteworthily, at   $s=1$ and $d=4$ we obtained  $m_h$ of the BH solutions that were derived in  \cite{Awad:2017tyz}.
\begin{figure}[ht]
\centering
\subfigure[~The value $m_h$ of the parameter $m$  in 4-dim.]{\label{fig:pot1}\includegraphics[scale=0.3]{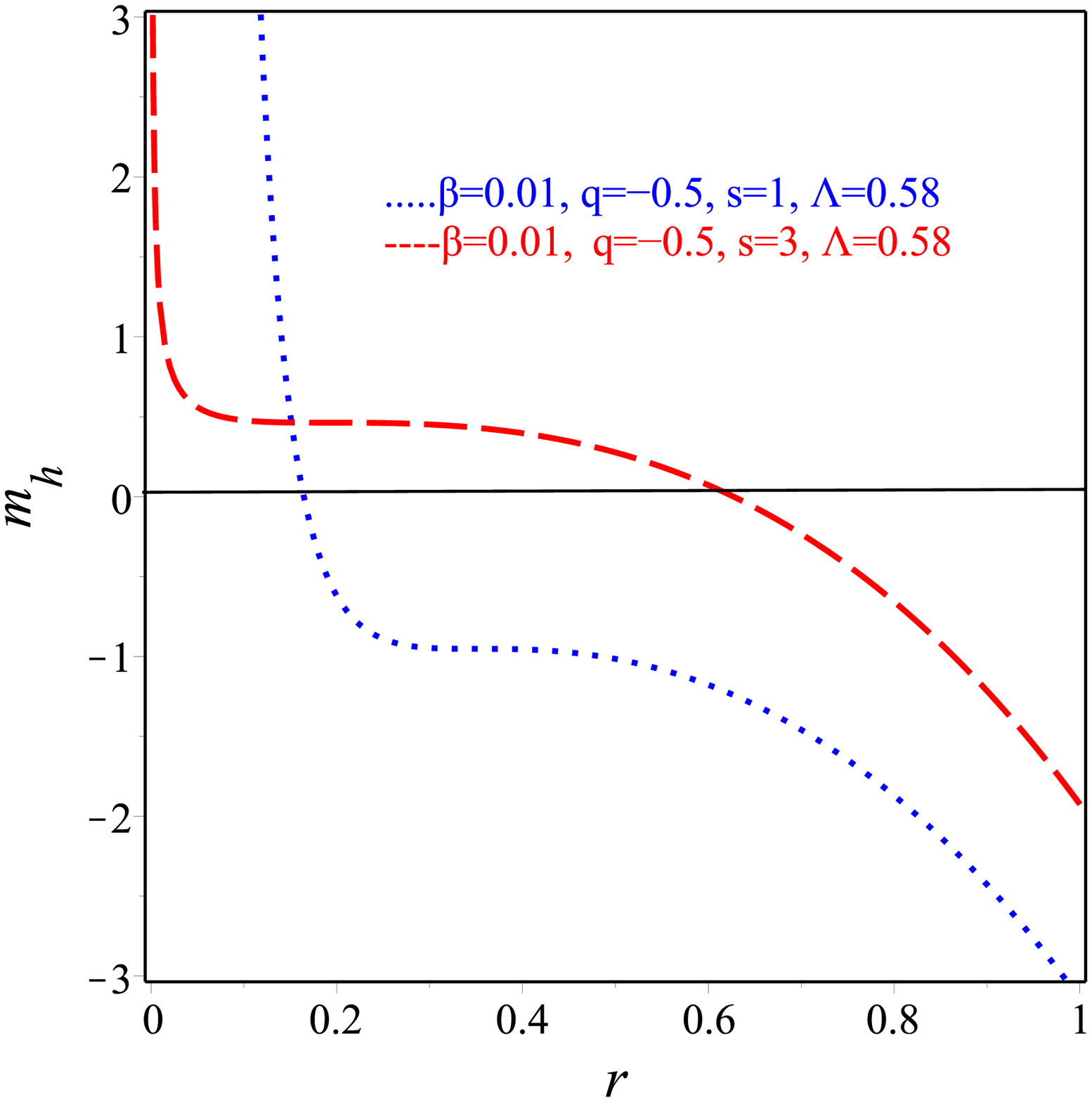}}
\subfigure[~The value $m_h$ of the parameter $m$  in 5-dim.]{\label{fig:pot1}\includegraphics[scale=0.3]{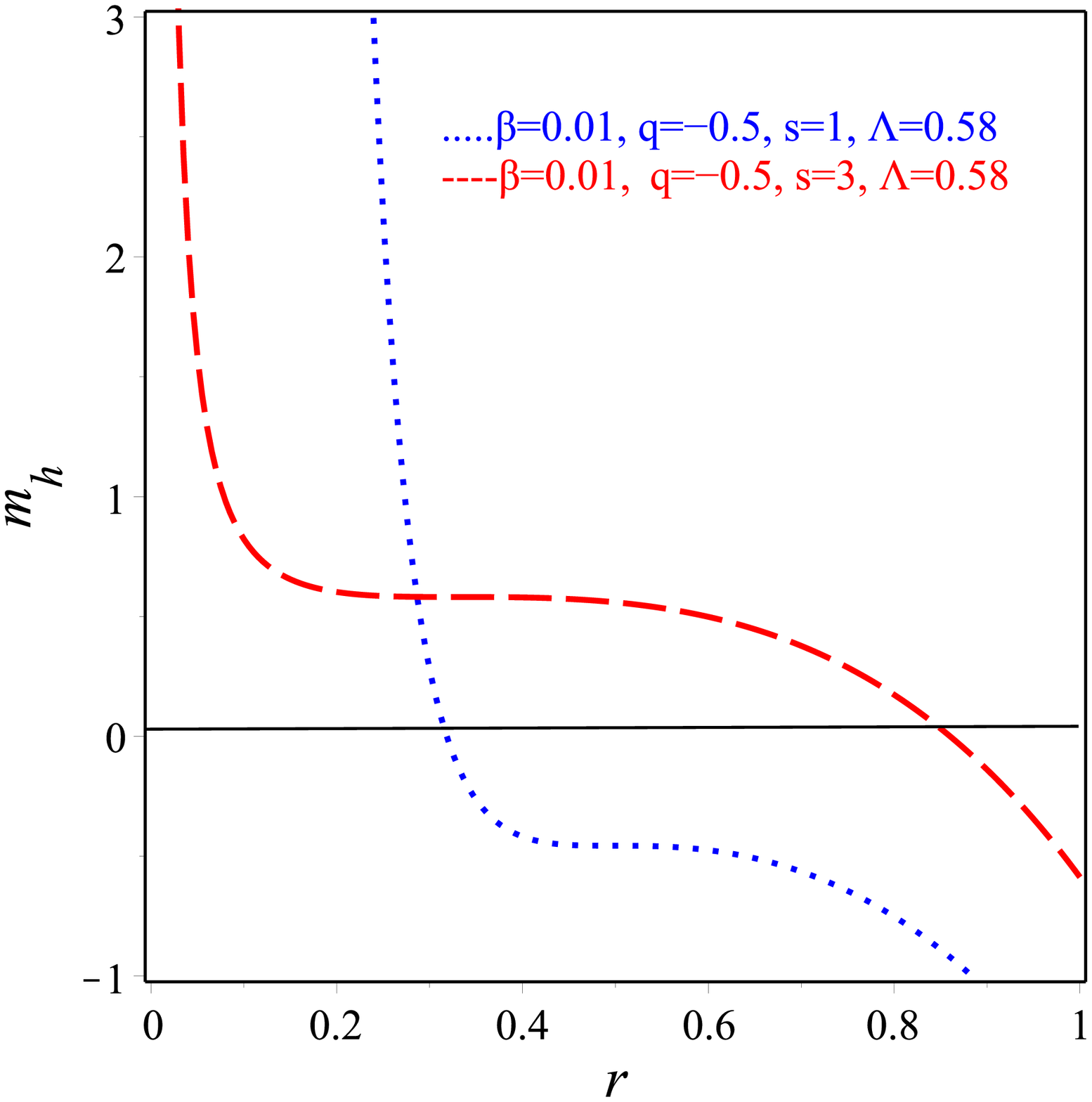}}
\caption{\it{{The value $m_h$ of the parameter $m$  that corresponds to the horizon
$r_h$, of solution (\ref{sol1}) of power law Maxwell-$f(T)$
gravity  in four and five dimensions,  for various values of  the power law parameter $s$. }}}
\label{Fig:2}
\end{figure}
{Eq. (\ref{hor1}) is plotted in Fig. \ref{Fig:2} when $d=4$ and $d=5$.}

\subsection{Energy of the black hole (\ref{sol1})}\label{S3energy}

In this subsection, the energy of the BH solution (\ref{sol1}) was calculated from Eq.   (\ref{qbbb}), and the necessary component of the super-potential  $S^{\mu \nu \rho}$  of the BH solution (\ref{sol1}) is given as follows:
\begin{equation}  \label{8}
S^{001}=\frac{g(r)}{r},
\end{equation}
where $g(r)$ is expressed by Eq. (\ref{sol1}).
The regularized expression of Eq. (\ref{conr}) takes the following form:
\begin{equation}  \label{reg} P^a:=\int_V d^{d-2}x  \left[e{S}^{a 0
0} f_T\right]-\int_V d^{d-2}x \left[e{S}^{a 0
0} f_T\right]_{AdS/dS},\end{equation}  where {\it AdS/dS} means calculations for pure AdS/dS space.  Using Eq. (\ref{reg}) in solution (\ref{sol1}),  we get
 \begin{eqnarray}\label{reg1}  && E=\frac{(d-2)m}{2}+\frac{2(d-3)^2(2s-1)2^{s-2}3^{s-1}q^{2s} }{(2s-3) r^{\frac{[d-1-2s]}{2s-1}}}+\frac{2^{s+1}(3s-1)(2s-1)(d-3)^4\sqrt{3(2s-1)\beta}q^{3s} }{3(2s-3)r^{\frac{(2d-5)}{2s-1}}}+\cdots\,.\end{eqnarray}
 Eq.  (\ref{reg1}) indicates that the modification of  $f(T)$ did not affect the mass term of the energy of standard TEGR \cite{Maluf:2002zc,Maluf:1995re}, although it affected the charge terms. The charge terms would aid the calculation of energy
starting from ${\mathcal{O}}\Bigg(\frac{1}{r}\Bigg)$, contrary to Reissner-Nordstr\"om
spacetime. This difference is due to the contribution of the function $g(r)$ given in
  Eq. (\ref{sol1}). Moreover, as already stated, $s$ must satisfy    $s\leq\frac{d-1}{2}$ so that the second term in Eq. (\ref{reg1}) could be finite at $r\rightarrow 0$.

 \section{Rotating BHs in the power law Maxwell-$f(T)$ gravity }\label{S5}

Further, the rotating BH solutions were derived to satisfy the field equations of the power-law $f(T)$ gravity. Thus, we utilized the static BH solutions that were derived in the previous section and applied the following transformations with $n$-rotation parameters, as follows:
\begin{equation} \label{t1}
\bar{\theta}_{i} =-\aleph~ {\theta_{i}}+\frac{ \omega_i}{\lambda^2}~t,\qquad \qquad \qquad
\bar{t}=
\Omega~ t-\sum\limits_{i=1}^{n}\omega_i~ \theta_i,
\end{equation}
where  $\omega_i$  are the   rotation parameters (their number is
$n= \lfloor(d - 1)/2\rfloor$ where  $\lfloor ... \rfloor$ represents the integer part), and  $\lambda$ is related to the parameter
$\Lambda_{eff}$ of the static solution (\ref{sol1})
through Eq. (\ref{36}):
\begin{eqnarray}\label{36}
\lambda=-\frac{(d-2)(d-1)}{2 \Lambda_{eff}}.
\end{eqnarray}
Additionally, the parameter
$\Omega$ is defined as follows:
\[\Omega:=\sqrt{1-\sum\limits_{j=1}^{{n}}\frac{\omega_j{}^2}{\lambda^2}}.\]
Applying the transformation (\ref{t1})
to the vielbeins (\ref{tetrad}) we obtained the following equation:
\begin{eqnarray}\label{tetrad1}
\nonumber \left({e^{i}}_{\mu}\right)=\left(
  \begin{array}{cccccccccccccc}
   \Omega\sqrt{N(r)} & 0 &  -\omega_1\sqrt{N(r)}&-\omega_2\sqrt{N(r)}\cdots &
-\omega_{n}\sqrt{N(r)}&0&0&\cdots&0
\\[5pt]
    0&\frac{1}{\sqrt{N(r)g(r)}} &0 &0\cdots &0&0&0&\cdots & 0\\[5pt]
          \frac{\omega_1r}{\lambda^2} &0 &-\Omega r&0  \cdots &0&0&0&\cdots & 0\\[5pt]
        \frac{a_2r}{\lambda^2} &0 &0  &-\Omega r\cdots & 0&0&0&\cdots & 0\\[5pt]
        \vdots & \vdots  &\vdots&\vdots&\vdots &\vdots&\vdots& \cdots & \vdots \\[5pt]
  \frac{ \omega_n\;r}{\lambda^2}  &  0 &0&0 \cdots & -\Omega r&0&0&\cdots & 0 \\[5pt]
   0 &  0  &0&0 \cdots &0&r&0&\cdots & 0\\[5pt]
     0 &  0  &0&0 \cdots &0&0&r&\cdots & 0\\[5pt]
       0 &  0 &0&0 \cdots &0&0&0&\cdots & r\\
  \end{array}
\right),&\\
\end{eqnarray}
   where $N(r)$ and $g(r)$ are given by Eq.  (\ref{sol1}). Hence, for the electromagnetic potential
(\ref{pot}), we obtained the following form:
\begin{equation}
\label{Rotpot}
\bar{\phi}(r)=-\phi(r)\left[\sum\limits_{j=1}^{n} \omega_j d\theta'_j-\Omega
dt'\right].
\end{equation}
Notably, although the transformation  (\ref{t1}) did not alter the local properties of the spacetime, it changed them globally, as reported in  \cite{Lemos:1994xp}, since it mixed the compact and noncompact coordinates. Thus, the vielbeins,  (\ref{tetrad}) and (\ref{tetrad1}) could only be locally mapped into each other \cite{Lemos:1994xp,Awad:2002cz}.

The metric, which corresponds to the \ {vielbein},  (\ref{tetrad1}) is written as follows:
\ba
\label{m1}
&&    ds^2=N(r)\left[\Omega d{t'}  -\sum\limits_{i=1}^{n}  \omega_{i}d{z}'
\right]^2-\frac{dr^2}{N(
r)g(r)}-\frac{r^2}{\lambda^2}\sum\limits_{i=1 }^{n}\left[\omega_{i}d{t}'+\Omega \lambda^2 d{\theta}'_i\right]^2-
\sum\limits_{k=1 }^{d-n-2}r^2 dz_k^2-\frac{r^2}{\lambda^2}\sum\limits_{i<j
}^{n}\left(\omega_{i}d{\theta}'_j-\omega_{j}d{\theta}'_i\right)^2,\nonumber\\
&&
\ea
where $0\leq r< \infty$, $-\infty < t < \infty$, $0 \leq \theta_{i}< 2\pi$, $i=1,2 \cdots
n$ and $-\infty < z_k < \infty$,   where  $d z_k^2$ is
the Euclidean metric on $(d-2)$ dimensions with $k = 1,,2\cdots d-3$. As mentioned earlier, the static configuration  (\ref{m2}) could be recovered as a special case of the aforementioned general metric if we choose to vanish the rotation parameters
$\omega_j$.

Finally, following the procedure in subsection \ref{S3energy} , the energy of the rotating charged AdS BH  (\ref{m1}), as calculated as follows:
\begin{equation}  \label{9}
 E= \frac{(d-2)[1+2(d-1)(d-2)\Lambda_{eff}\beta][\Lambda_{eff}{}^2 \omega_j\sum\limits_{j=1}^{{n}}\omega_j+3\Omega^2]m }{12(d-3)G_d}
+\cdots.
\end{equation}
{Equation (\ref{9}) is the energy of spacetime (\ref{tetrad1}) which contains the rotation parameters $\omega_j$ and when these parameters are vanishing we get the value of energy given by Eq. (\ref{reg1}) provided the use of Eq. (\ref{lam}).}
\section{Thermodynamics of the derived black holes}\label{S7}
\begin{figure}[ht]
\centering
\subfigure[~The value of the entropy, $S_h$  in 4-dim.]{\label{fig:Ent1}\includegraphics[scale=0.3]{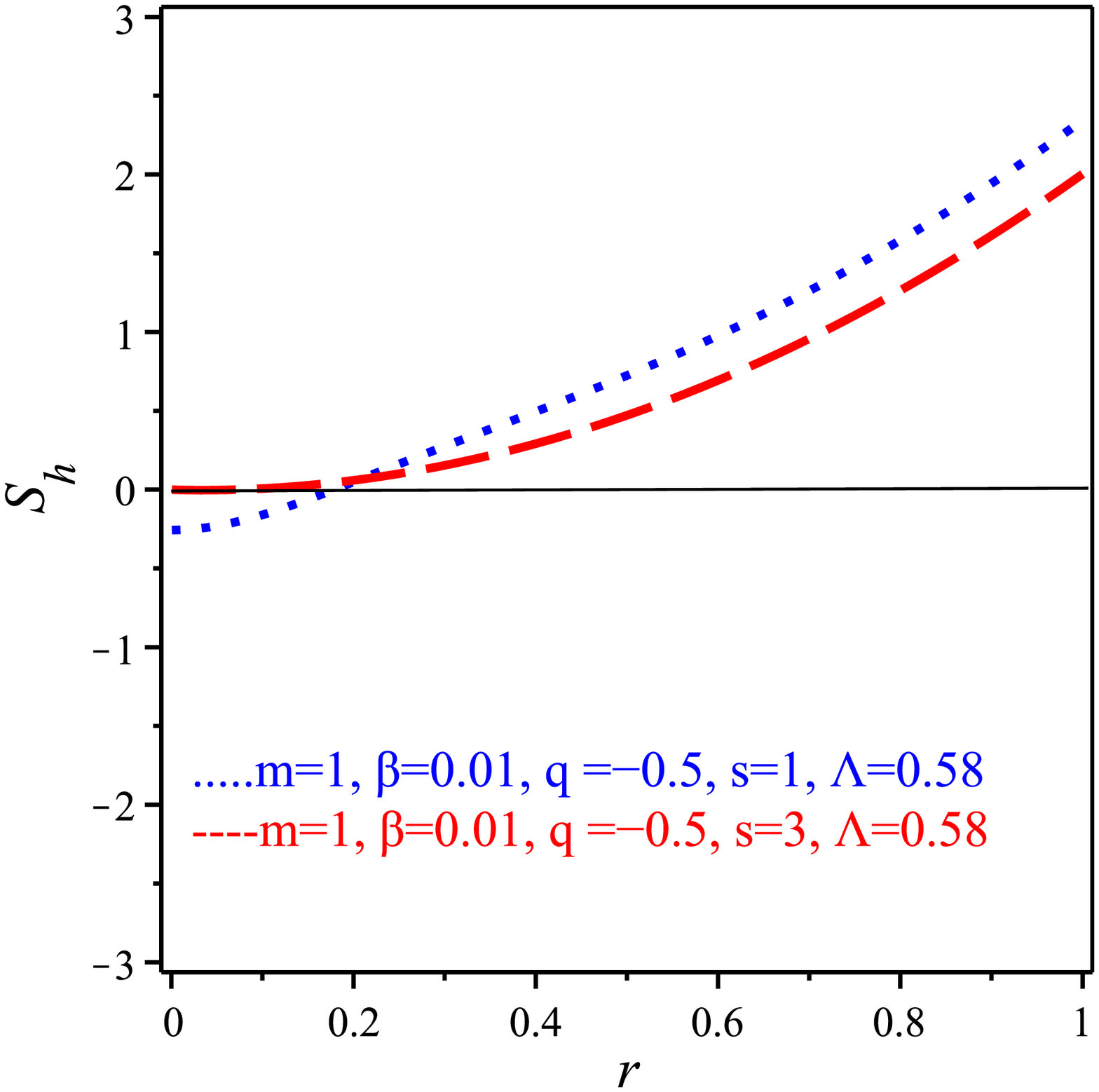}}
\subfigure[~The value of the entropy, $S_h$ in 5-dim.]{\label{fig:Ent2}\includegraphics[scale=0.3]{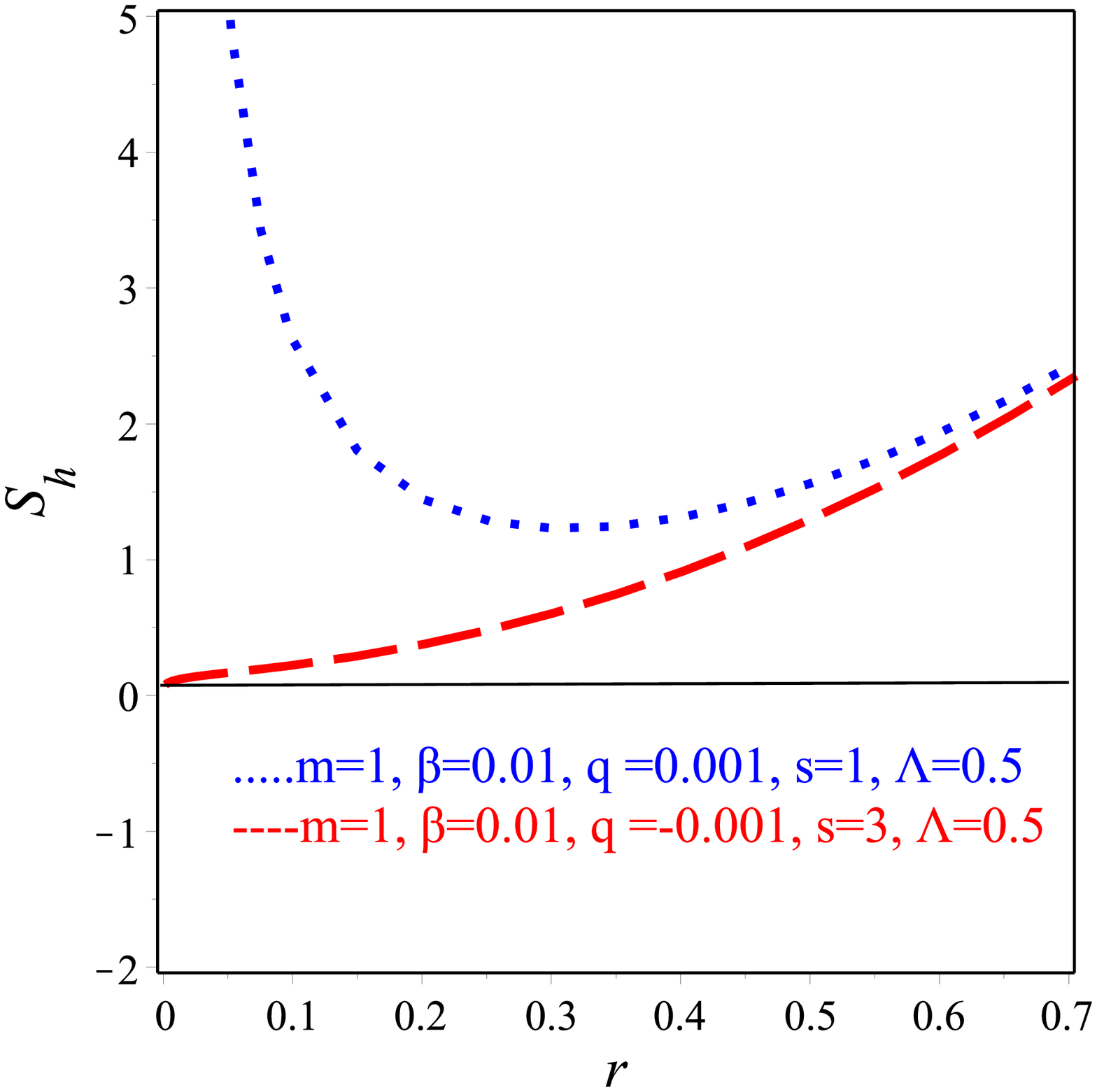}}
\caption{\it{{The value of the entropy, $S_h$,  that corresponds to the horizon
$r_h$, of solution (\ref{sol1}) of power law Maxwell-$f(T)$
gravity  in four and five dimensions,  for various values of  the power law parameter $s$.  }}}
\label{Fig:3}
\end{figure}

To study different thermodynamical properties \cite{Hunter:1998qe,Hawking:1998ct,Bekenstein:1972tm,Bekenstein:1973ur,Gibbons:1977mu}   of BH solution (\ref{sol1})  we start  by defining the Bekenstein--Hawking entropy  of $f(T)$ as reported in  \cite{Karami:2012fu,Bamba:2012vg,2015EPJP..130..124N,2015JPSJ...84d4006N}
\begin{equation}\label{ent}
S(r_h)=\frac{1}{4}Af_T=\pi r_h{}^2(1-2\beta T),
\end{equation}
where $A$  is the area of the event horizon and $T$ is the scalar torsion, which is given by Eq.  (\ref{tors}). {We plotted the entropy relation in Fig. \ref{Fig:3} for $d=4$ and $d=5$. The figure \ref{Fig:3} \subref{fig:Ent1} revealed that we obtained a negative entropy in the region when  $r<r_{dg}$. At $r>r_{dg}$, we obtained a positive value. Eq. (\ref{ent}) revealed that the entropy was not proportional to the area because of the appearance of $\beta$. However, when $d=5$ we always obtain a positive entropy as \ref{Fig:3} \subref{fig:Ent2} shows.} The heat capacity,  $C_h$ which is valuable for examining the stability of BH, was defined thermodynamically according to its sign. $C_h$ is defined as reported in  \cite{Nouicer:2007pu,DK11,Chamblin:1999tk,Nashed:2018qag}
\begin{equation}\label{m55}
C_h=\frac{\partial m}{\partial T}\equiv \frac{\partial m}{\partial r_h} \left(\frac{\partial r_h}{\partial T}\right).
\end{equation}
Therefore, if the heat capacity  is is positive, $C_{h} > 0$, BH would thermodynamically stable. Conversely, at  ($C_h< 0$), BH would be thermodynamically unstable.

\begin{figure}[ht]
\centering
\subfigure[~The value of the Hawking temperature, $T_h$  in 4-dim.]{\label{fig:pot1}\includegraphics[scale=0.3]{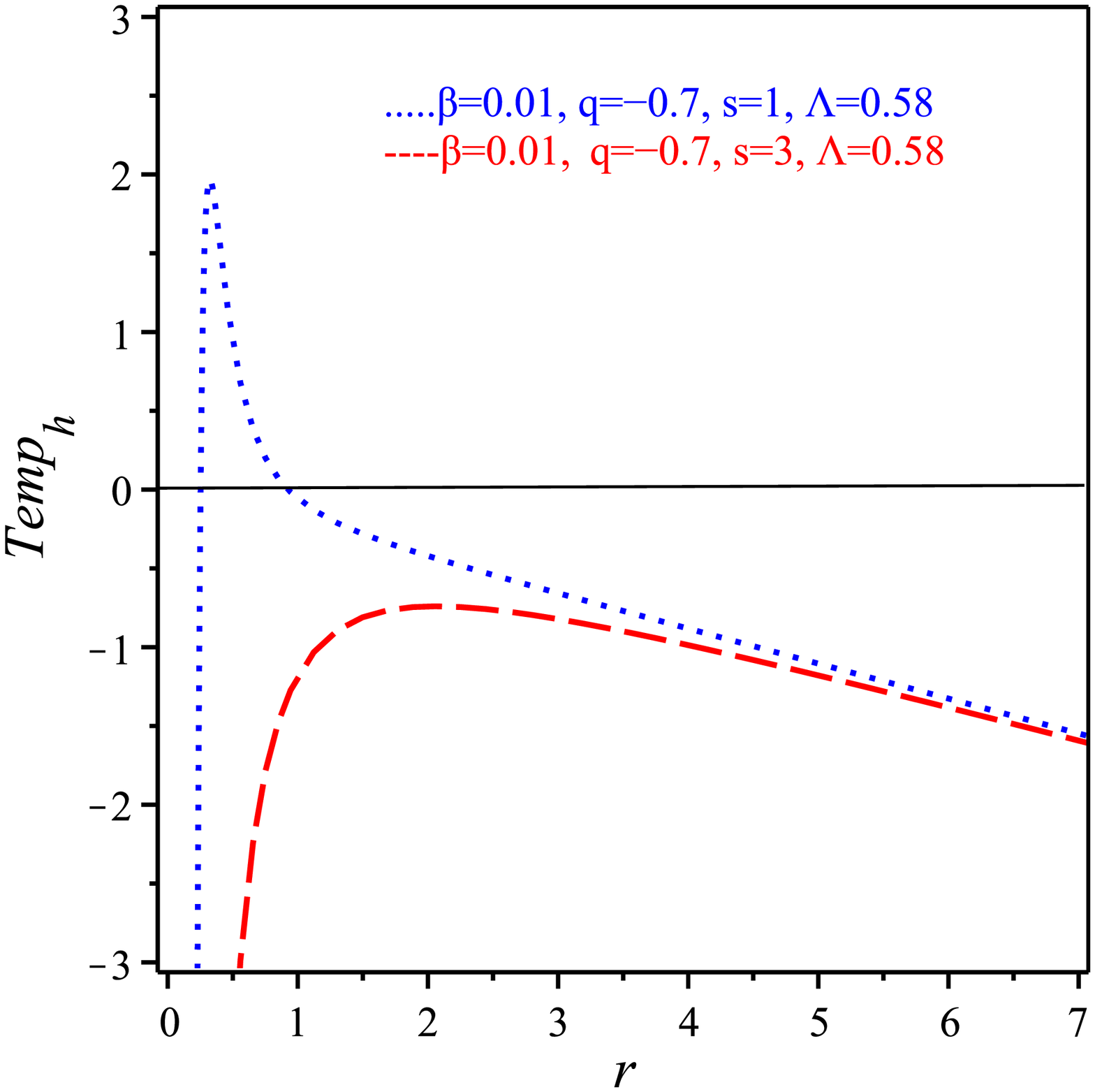}}
\subfigure[~The value of the Hawking temperature, $T_h$ in 5-dim.]{\label{fig:pot1}\includegraphics[scale=0.3]{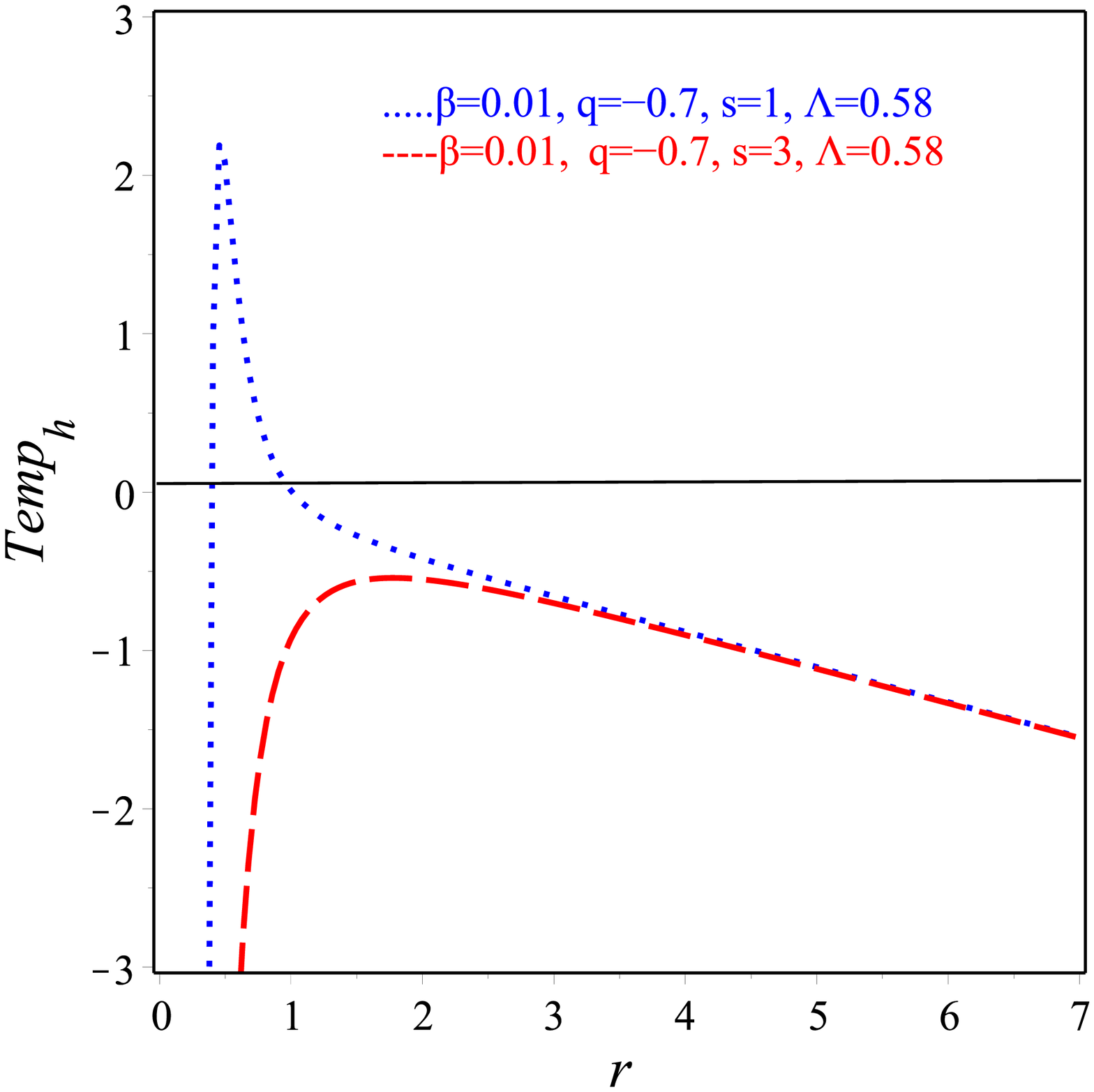}}
\caption{\it{{The value of the Hawking temperature, $T_h$,  that corresponds to the horizon
$r_h$, of solution (\ref{sol1}) of power law Maxwell-$f(T)$
gravity  in four and five  dimensions,  for various values of  the power law parameter $s$.  }}}
\label{Fig:4}
\end{figure}

\begin{figure}[ht]
\centering
\subfigure[~The value of the heat capacity, $C_h$ in 4-dim.]{\label{fig:hc1}\includegraphics[scale=0.3]{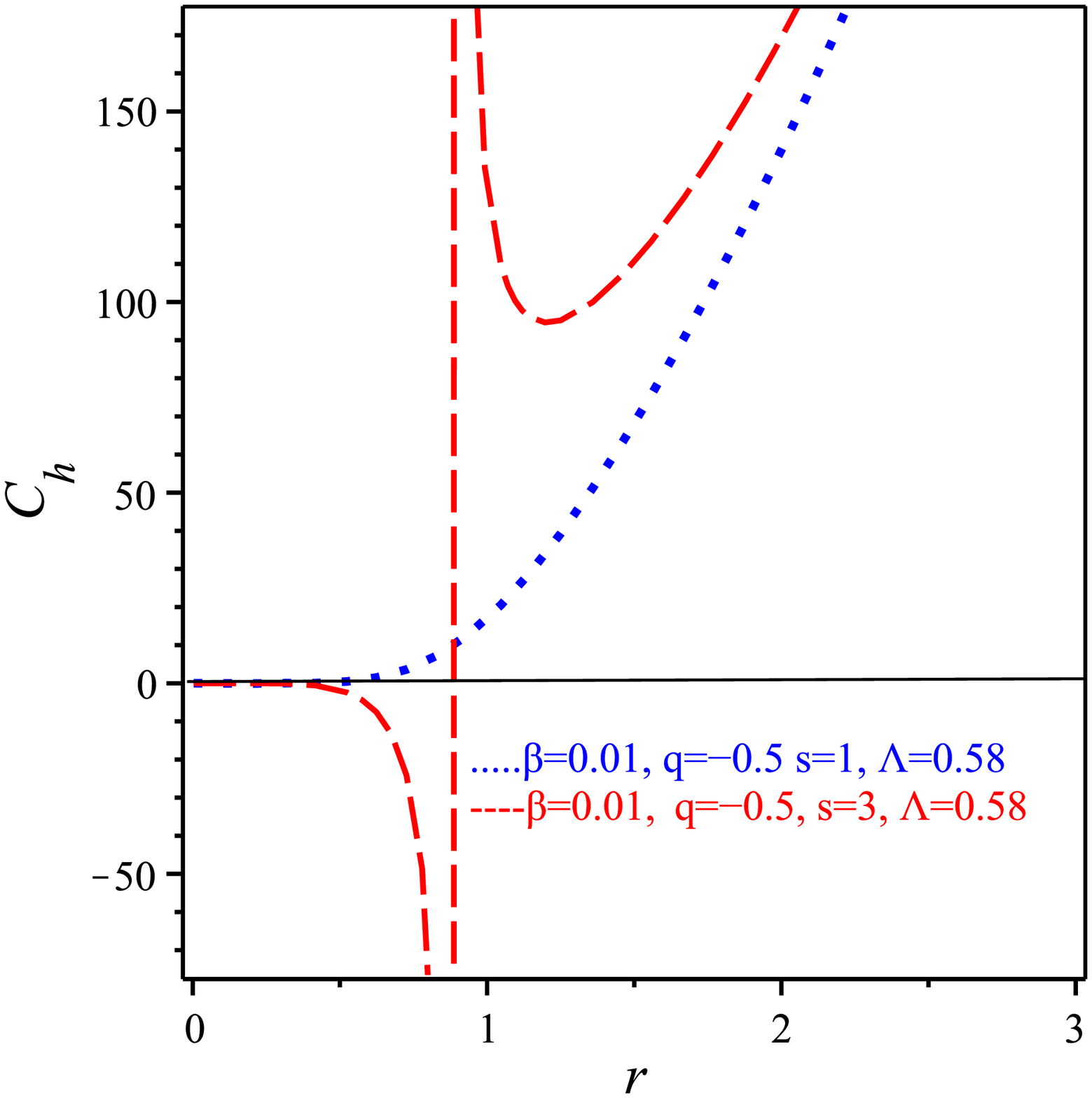}}
\subfigure[~The value of the heat capacity, $C_h$ in 5-dim.]{\label{fig:hc2}\includegraphics[scale=0.3]{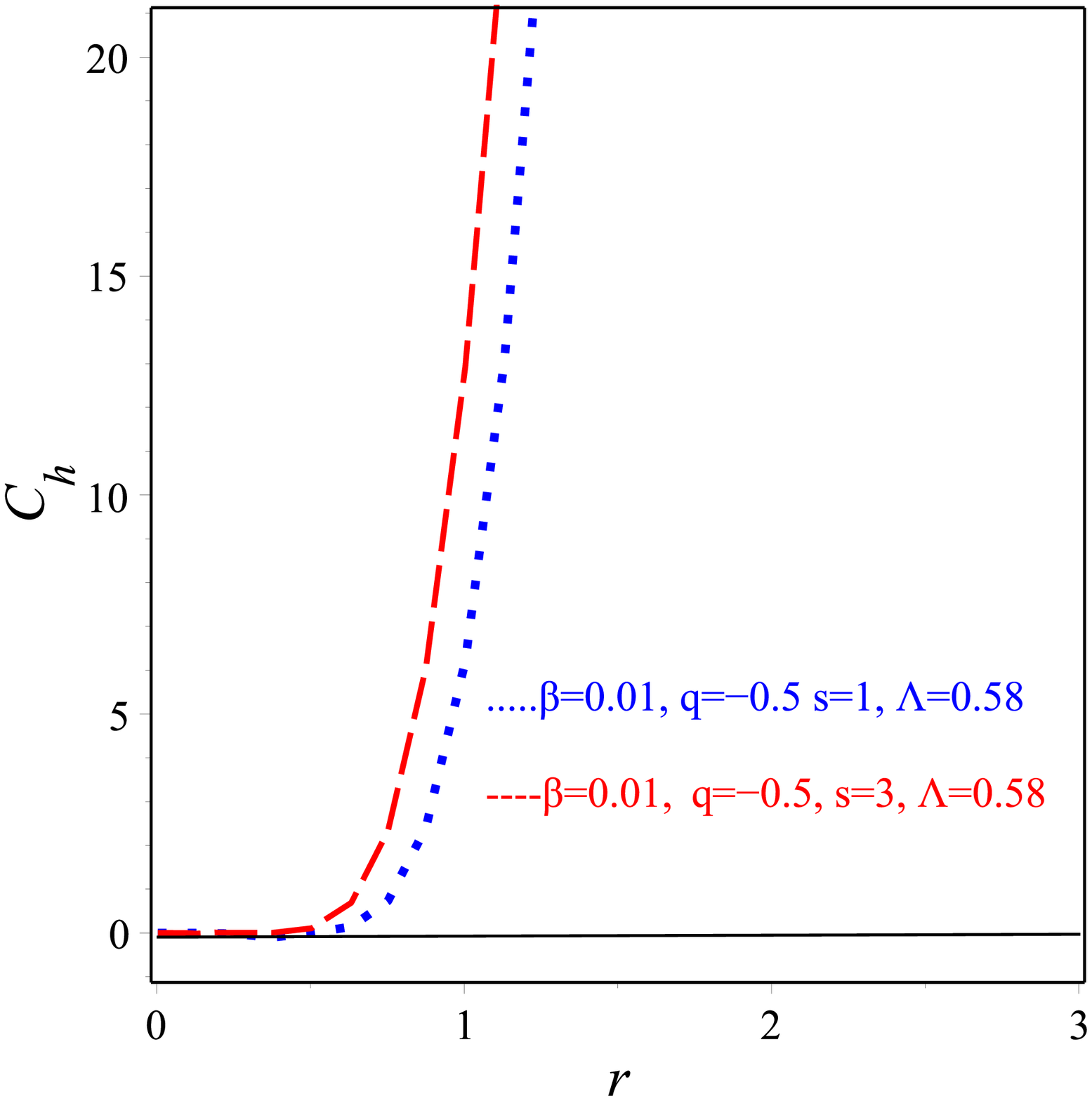}}
\caption{\it{{The value of the heat capacity, $C_h$,  that corresponds to the horizon
$r_h$, of solution (\ref{sol1}) of power law Maxwell-$f(T)$
gravity  in four and five dimensions,  for various values of  the power law parameter $s$. }}}
\label{Fig:5}
\end{figure}

To calculate Eq. (\ref{m55}) we calculate the mass of BH mass in  $r_h$ as obtained in Eq.  (\ref{hor1}).  The Hawking temperature of BHs can be defined as reported in \cite{Hawking:1974sw}
\begin{equation}
T = \frac{\kappa}{2\pi}, \qquad \textmd{where} \quad \kappa \quad \textmd{ is the surface gravity which is defined as, } \qquad \qquad \kappa= \frac{N'(r_h)}{2}.
\end{equation}
The Hawking temperatures which is associated with the BH solution  (\ref{sol1}), is expressed as follows:
\begin{equation} \label{m44}
T_h= \frac{(d-1)(d-2)r_h{}^2-24\;2^s(2s-1)q^{3s}\beta \sqrt{2^s\; (-1)^{^{s+1}}\; 3\;\beta(2s-1)}\;r^{-\beta_2}-36\beta q^{2s}(2s-1)2^s\; r^{-\beta_1}}{96r_h \beta (d-2)},
\end{equation}
where ${T_h}$ is the Hawking temperature at the inner horizon. {We plotted  $T_h$ in Fig. \ref{Fig:4} for $d=4$ and $d=5$.}
Further, we calculated  $C_{h}$ after substituting Eq. (\ref{hor1}) and (\ref{m44}) into   (\ref{m55}) as follows:
\begin{eqnarray} \label{m66}
&&C_h=\Big(\frac{4\pi \sqrt{2s-1}r_h{}^{d-2}}{A(24\;2^s\;q^{3s}\;(2s-1)
\sqrt{2^s\;(-1)^{^{s+1}}\; 3\;\beta}(1+3sd-8s)\beta r_h{}^{-\beta_2}+\sqrt{2s-1}\{r_h{}^2(d-1)(d-2)-36\;q^{2s}\;2^s\;\beta\;r_h{}^{-\beta_1}\})}\Big)\nonumber\\
&&\times\Big(24\;q^{3s}\;2^s\;\sqrt{2^s\; (-1)^{^{s+1}}\; 3\;\beta(2s-1)}(2s-1)[(2+3sd-10s)-(2s-1)(d-3)]\beta\;r_h{}^{-\beta_2}\nonumber\\
&&-A[r_h{}^2(d-1)(d-2)-36\;q^{2s}\;2^s\;(2s-1)\;\beta\;r_h{}^{-\beta_1}]\Big),
\end{eqnarray}
{It was challenging to extract any information from Eq.  (\ref{m66}), therefore, we plotted it in 4-dimension as  illustrated in Fig. \ref{Fig:5}, for particular values of the BH parameters. As figure \ref{Fig:5} \subref{fig:hc2} shows that we have always positive heat which means  a stable BH in 5-dimension.}

Although  Fig. \ref{Fig:5} indicates that  $C_{h}$   of the linear case was always positive, it was negative in the nonlinear case and diverged at some critical values, $r_h < r_{min}$ a, thus exhibiting a positive value. These imply that BH of the linear Maxwell--$f(T)$ exhibited local stability, whereas the nonlinear case exhibited local stability only when  $r_h >r_{min}$, otherwise, it did not.

\begin{figure}[ht]
\centering
\subfigure[~The value of the free energy, $G_h$ in 4-dim.]{\label{fig:pot1}\includegraphics[scale=0.3]{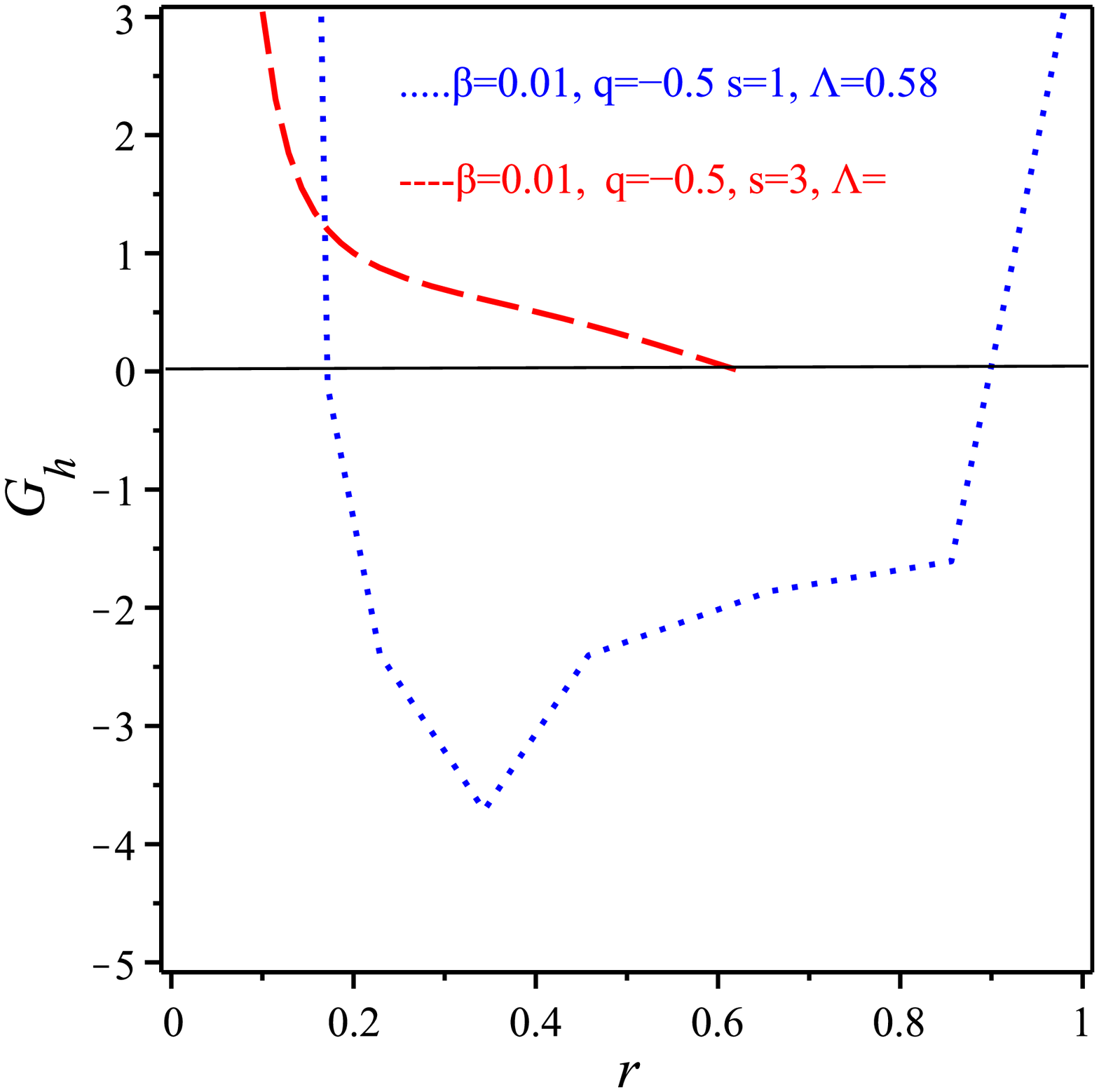}}
\subfigure[~The value of the free energy, $G_h$ in 5-dim.]{\label{fig:pot1}\includegraphics[scale=0.3]{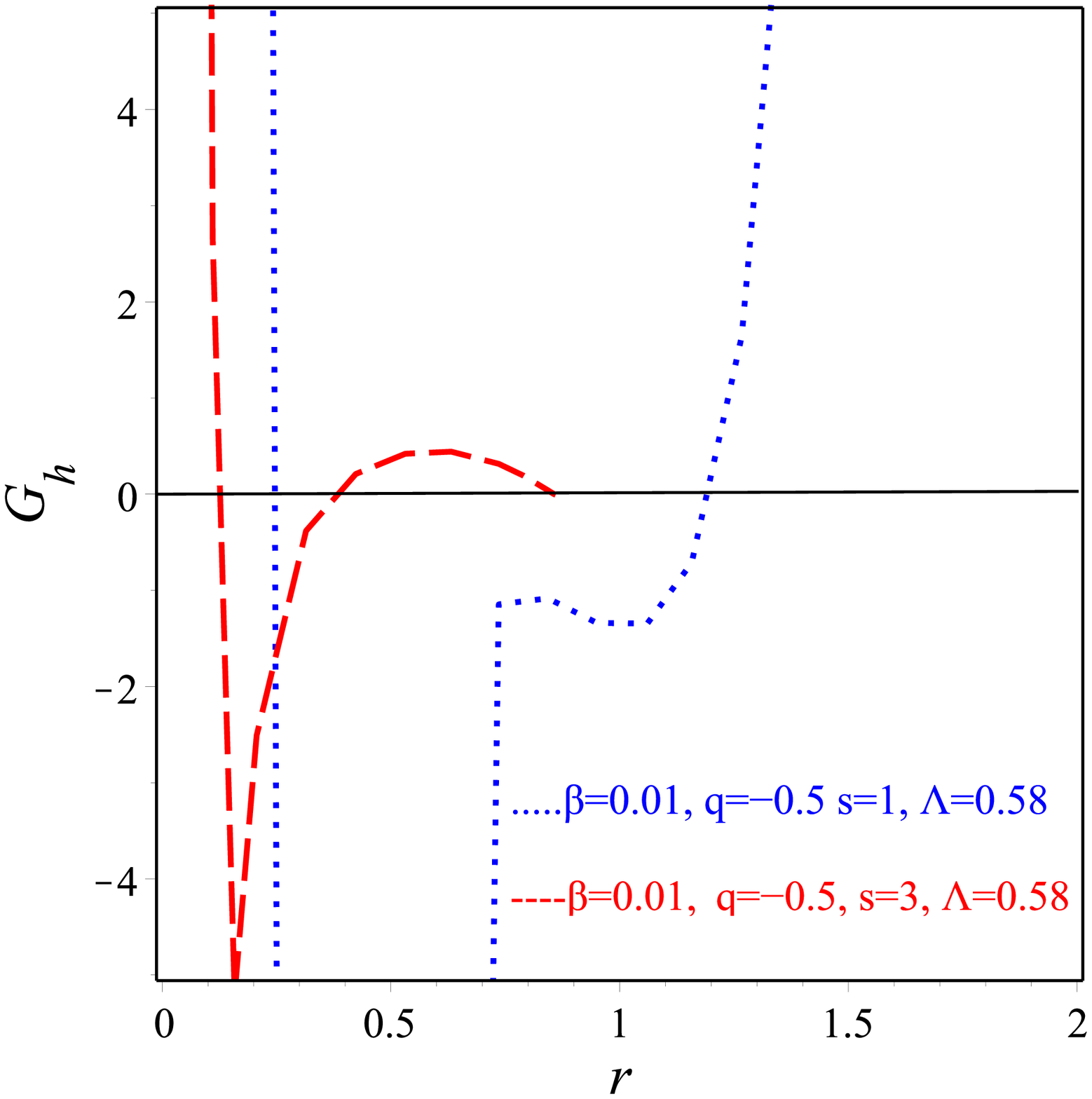}}
\caption{\it{{The value of the free energy, $G_h$,  that corresponds to the horizon
$r_h$, of solution (\ref{sol1}) of power law Maxwell-$f(T)$
gravity  in four and five  dimensions,  for various values of  the power law parameter $s$.  }}}
\label{Fig:6}
\end{figure}

The free energy in the grand canonical ensemble, which is called the Gibbs free energy, is defined as reported in  \cite{Nashed:2018qag}:
\begin{equation} \label{enr}
G(r_h)=M(r_h)-T(r_h)S(r_h),
\end{equation}
where $M(r_h)$, $T(r_h)$ and $S(r_h)$ are the mass of the BH, the temperature and entropy  at the event horizon, respectively.   Inserting  Eqs. (\ref{hor1}), (\ref{ent}) and (\ref{m44}) into (\ref{enr}) ) affords a lengthy expression. Here we just demonstrated the behavior of the free energy in 4-dimension, as illustrated in  Fig. \ref{Fig:6}.

{Notably, when $d=4$,  the linear Maxwell field had a positive value, followed by a negative one in the range of  $0.2<r<0.8$ and a positive one again  \cite{Altamirano:2014tva}. However, the nonlinear Maxwell field was always a positive quantity, and this indicates that the nonlinear Maxwell field was always locally stable. Same discussion can be applied for $d=5$ but in that case the linear one has a positive value then a negative value in the range $0.2<r<0.4$ and the always positive value. }
\section{Conclusions and Discussions}\label{S8}
We investigated the effect of the nonlinear power law of electrodynamics in the context of the modified TEGR theory,  {$f(T)$-gravity}. To do this, we applied the flat horizon spacetime in a diverse dimension and applied it to a specific form of  $f(T)=T_0+\alpha T-\beta T^2$, {where $T_0$ and $\alpha$ are constants and  $\beta$ is a dimensional parameter.} The obtained nonlinear second-order differential equations were solved in an exact approach. {The BH solution of these differential equations was characterized by two integration constants, in addition to the nonlinear parameter that describes the nonlinearity of the Maxwell field equation. The two constants were explained to represent the mass and charge of BH.} This BH solution is a generalization to that, which was presented in   \cite{Awad:2017tyz} owing to the existence of the nonlinear parameter. When it was equal to 1, we returned to BH that was discussed in \cite{Awad:2017tyz}.

To investigate the physical properties of this generalized BH, we calculated the scalar invariants (the ones that are related to the curvature and those that are related to the torsion) and demonstrated that the singularities of this BH were much softer than those of  \cite{Awad:2017tyz} owing to the contribution of the nonlinear electromagnetic parameter. Moreover, we demonstrated that the singularities of our BH were much milder than those of GR BH. This result is considered to be the main merit of this study, {in addition to the fact that the calculations of the energy confirmed that $s$ affects the asymptotic behavior of the charged terms as shown in Eq. (\ref{reg1})}.


To explore our BH in detail, we applied a coordinate to create an exact rotating BH with nonlinear electrodynamics in the  $f(T)$ frame. The features of this rotating BH are that it possessed $d$-dimensional rotating parameters, and could easily return to the nonrotating BH if all the rotation parameters were set to zero. Notably, all the features of the singularities of the nonrotating BH were present in the rotating one.

 Finally, we calculated some thermodynamic quantities of the nonrotating case and revealed that the entropy was not proportional to the area. The entropy was non-proportional to the area because our BH possessed a non-vanishing value of $T$. Our calculations revealed that the entropy might have a negative value when  $r_h<r_{dg}$ otherwise, it would have a positive value. A negative entropy has been obtained and explained \cite{Cvetic:2001bk,Nojiri:2001fa,Nojiri:2002qn,Nojiri:2017kex,Clunan:2004tb,Nojiri:2001ae}. The entropy might be negative because the dimensional parameter, $\alpha$, had entered an unpermitted phase.

Further, we computed the temperature of BH and demonstrated that its value was negative. This value mainly accounts for the structure of ultracold BH  \cite{Davies:1978mf}. The computations of the thermodynamics in the context of the {$f(T)$-gravity} for a non-trivial BH, which possessed a non-vanishing value of $T$, achieved a limit for validation. This statement required further study, which would be conducted in another work.

\subsection*{Acknowledgments}
The authors acknowledge the anonymous referee for improving the presentation of the manuscript. G.N. would like to thank TUSUR for visiting fellowship. The work of KB has partially been supported by the JSPS KAKENHI Grant Number JP21K03547.

%

\end{document}